\documentclass[
reprint,
amsmath,amssymb,aps, showkeys,
physrev,
]{revtex4-2}

\usepackage{graphicx}
\usepackage{bm}

\usepackage[colorlinks=true,citecolor=blue,linkcolor=blue,breaklinks=true,urlcolor=blue]{hyperref}
\usepackage{orcidlink}
\usepackage{arydshln}  
\usepackage{booktabs}  

\begin{document}

\preprint{APS/123-QED}

\title{A Tentative Double Excess in the Gamma-Ray Spectrum of the Fermi Blazar 4FGL J0604.9-0000}

\author{Shi-Ju Kang\orcidlink{0000-0002-9071-5469}}
\email{kangshiju@alumni.hust.edu.cn}
\affiliation{School of Physics and Electrical Engineering,  Liupanshui Normal University,  Liupanshui, Guizhou, 553004, People's Republic of China}

\date{\today}

\begin{abstract}

Based on 17.75 years of Fermi-LAT observations, we report a tentative double excess in the gamma-ray spectrum of the blazar 4FGL J0604.9-0000, with peak energies at approximately 1.5 GeV and 11 GeV. The two excesses are modeled as double Gaussian components. For the lower-energy excess, the best-fit centroid is $E_1 = 1.59\pm0.07$ GeV with a width fixed to the instrumental resolution ($\sigma_1 = 0.145$ GeV). For the higher-energy excess, the centroid is $E_2 = 11.15\pm0.61$ GeV and the width is constrained to $\sigma_2 = 1.185$ GeV (about 10\% of the peak energy), which is comparable to the Fermi-LAT energy resolution at that energy. The two features have local significances of $2.6\sigma$ and $3.7\sigma$ (3 dof), respectively. A joint likelihood analysis yields a combined test statistic of TS $\simeq 27$, corresponding to a local significance of approximately $4.3\sigma$ (4 dof, without correction for trials and the look-elsewhere effect) or about $4.8\sigma$ (2 dof). To our knowledge, no other active galactic nucleus has been reported to show a similar double-excess candidate. The observed energy ratio of $\sim$1:7 is difficult to explain with standard astrophysical emission processes. However, the energies and their ratio are consistent with dark matter annihilation (e.g., $\chi\chi\to\gamma\gamma$ and $\chi\chi\to\gamma\gamma'$) for a particle mass near 11 GeV, making this source a promising target for follow-up observations with next-generation gamma-ray telescopes.

\end{abstract}

                              
\maketitle

\tableofcontents


\section{\label{sec:intro}Introduction} 

Blazars are a subclass of active galactic nuclei (AGN) whose relativistic jets are oriented very close to our line of sight, leading to extreme Doppler boosting, rapid variability, and dominant non-thermal emission across the entire electromagnetic spectrum \cite{1995PASP..107..803U,2019ARA&A..57..467B}. 
Their spectral energy distributions typically exhibit a double-hump structure: the low- energy peak (radio to X-rays) is widely attributed to synchrotron radiation from relativistic electrons in the jet, while the high-energy peak (MeV to TeV) is commonly explained by inverse Compton  scattering of soft photons by the same electron population, either internally (synchrotron self- Compton) or externally (external Compton) \cite{1992ApJ...397L...5M,1994ApJ...421..153S} or hadronic models involving proton-synchrotron or photo-pion processes \cite{1993A&A...269...67M,2001APh....15..121M,2003APh....18..593M,2013ApJ...768...54B,2019Galax...7...20B,2019NewAR..8701541H}.

The Fermi Large Area Telescope (LAT, \cite{2009ApJ...697.1071A}) has revolutionized the GeV gamma-ray view of blazars, revealing thousands of sources with diverse spectral behaviours \cite{2020ApJS..247...33A,2020ApJ...892..105A,2023arXiv230712546B,2026arXiv260222148B}.
In recent years, detailed Fermi-LAT observations have uncovered increasingly complex spectral features in blazars that challenge the conventional gamma-ray emission paradigm. These features include spectral breaks, hard-inverted spectra, and narrow line-like or bump-like structures in the gamma-ray band \cite{2009ApJ...699..817A, 2019A&A...623A...2A,2025ApJ...988..268L,2025ApJ...991L...8P,2025A&A...703A.162D,2021JCAP...08..007Z,2026arXiv260400579K}. Such features may hint at additional particle acceleration components, multiple emission zones, or even exotic physics beyond the standard astrophysical framework.

A particularly powerful signature, often dubbed a `smoking-gun' for dark matter (DM), is the detection of a monochromatic gamma-ray line arising from the annihilation or decay of DM particles \cite{1998APh.....9..137B,2012PDU.....1..194B,2018RvMP...90d5002B}. 
For instance, the process $\chi\chi\to\gamma\gamma$ would produce a line at $E_\gamma = m_\chi$, while $\chi\chi\to\gamma\gamma'$ (with $\gamma'$ a dark photon) could generate a second line at a lower energy. Such sharp spectral features are essentially free of conventional astrophysical backgrounds. 
Over the past decade, numerous line searches \cite{2010PhRvL.104i1302A} have been performed toward the Galactic center \cite{2012JCAP...08..007W,2015PhRvD..91l2002A,2016PDU....12....1D,2017ApJ...840...43A}, dwarf spheroidal galaxies \cite{2014PhRvD..89d2001A,2015PhRvL.115w1301A,2017ApJ...834..110A,2018JCAP...11..037A,2026JCAP...03..035A}, and galaxy clusters \cite{2016PhRvD..93j3525L,2021ApJ...920....1S,2024arXiv240711737F,2025PDU....4901966M}, but no conclusive signal has been established, with the strongest upper limits now approaching the thermal relic cross section for low DM masses.

Recently, the Galactic center has also exhibited a broad gamma-ray excess in the 1–3 GeV range, the origin of which remains debated \cite{2016PDU....12....1D}. Intriguingly, several independent works have reported tentative line-like features in individual blazars and radio galaxies \cite{2021JCAP...08..007Z,2026arXiv260400579K}. These findings motivate a systematic search for narrow gamma- ray features in AGN, where DM annihilation could be enhanced by a dense DM spike around the central supermassive black hole \cite{1999PhRvL..83.1719G,2001PhRvD..64d3504U,2002PhRvL..88s1301M,2022PhRvL.128v1104W,2024arXiv240601705C}.

In this work, we present a detailed analysis of 17.75 years of Fermi- LAT data toward the blazar 4FGL~J0604.9-0000. We discover two distinct excesses at approximately 1.5 GeV and 11 GeV that are well described as narrow Gaussian components. 
In Section~2 we describe the Fermi- LAT data reduction and analysis methods. Section~3 presents the spectral and temporal properties of the double excess, including systematic checks. In Section~4 we discuss possible astrophysical interpretations and compare with DM models. Section~5 summarizes our conclusions and outlines prospects for future observations.

\section{\label{sec:analysis}Data Analysis}

We analyzed $\sim$17.75 years of Fermi-LAT Pass 8 (P8R3) data from 2008 August 4 to 2026 May 4, targeting the blazar 4FGL~J0604.9-0000. Data reduction was performed using the Fermi Science Tools v2.2.0. Events within a $15^{\circ}$ region of interest (ROI) centered on the source were selected in the energy range 100~MeV–1~TeV, with standard quality cuts and zenith angle $<90^{\circ}$ to reduce Earth-limb contamination, using the P8R3\_SOURCE\_V3 instrument response functions. The background model included all 4FGL-DR4 sources \citep{2020ApJS..247...33A,2022ApJS..260...53A,2026arXiv260222148B} within the ROI, the Galactic diffuse model \texttt{gll\_iem\_v07.fits}, and the isotropic component \texttt{iso\_P8R3\_SOURCE\_V3\_v1.txt}; normalization of diffuse components and spectral parameters of sources within $5^{\circ}$ of the target were left free. We performed binned maximum likelihood fits comparing a null hypothesis (single log-parabola continuum) and a signal hypothesis (log-parabola plus two Gaussian components representing the excesses at $\sim$1.5 GeV and $\sim$11 GeV, with free amplitudes, centroids $E_{1,2}$, and widths $\sigma_{1,2}$) to obtain the maximum likelihood values $\mathcal{L}_{\rm null}$ and $\mathcal{L}_{\rm signal}$. The test statistic $\mathrm{TS}=2(\mathcal{L}_{\mathrm{signal}}-\mathcal{L}_{\mathrm{null}})$ follows a $\chi^2$ distribution with six (or fewer) degrees of freedom depending on the number of free Gaussian parameters; local significance is $\sigma=\sqrt{\mathrm{TS}}$.

\section{\label{sec:results}Results}

4FGL~J0604.9-0000 is a BCU-type blazar at $(l,b)=(207.58^{\circ},-10.33^{\circ})$ with a $14.0\sigma$ detection in the 4FGL-DR4 catalog \citep{2020ApJS..247...33A}. Its $\gamma$-ray spectrum is well described by a log-parabola ($\alpha=2.01\pm0.13$, $\beta=0.12\pm0.07$), and no spectroscopic redshift is available.

\begin{figure*}[htbp!]
    \centering
    \includegraphics[width=0.32\linewidth]{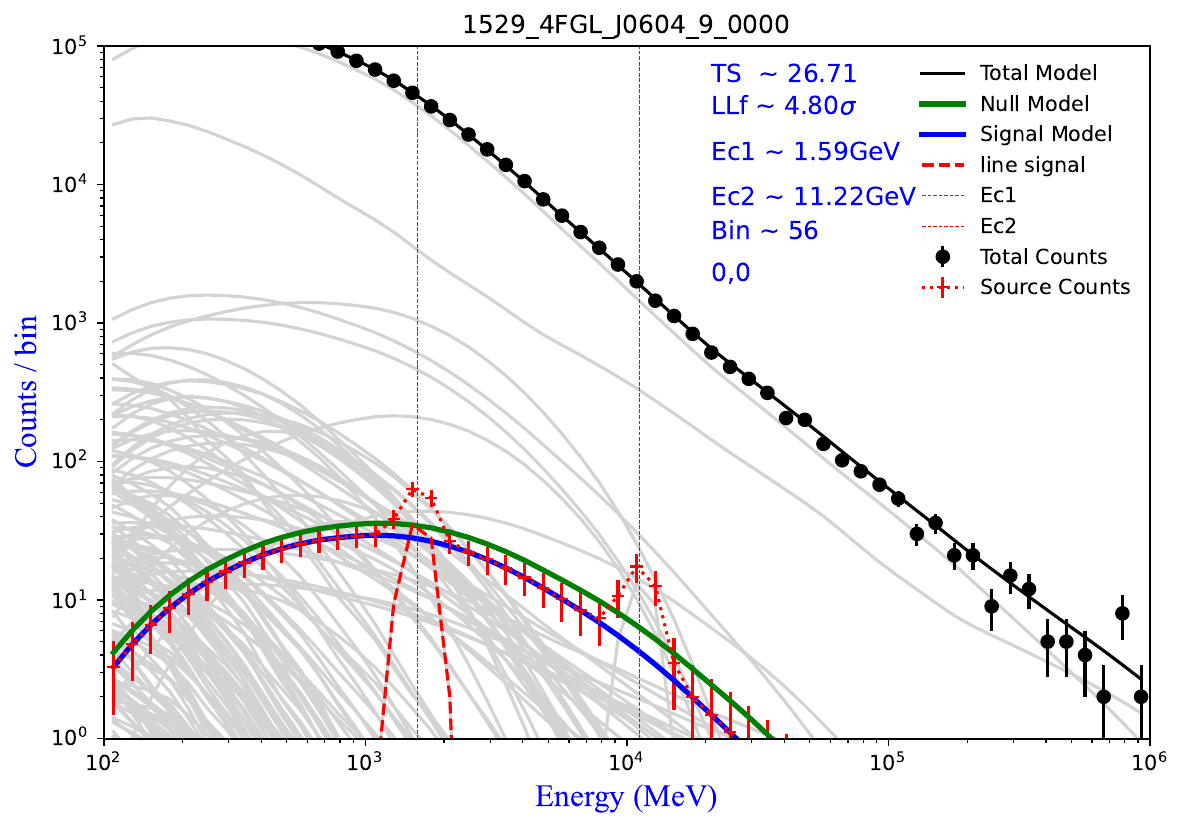}
    \includegraphics[width=0.32\linewidth]{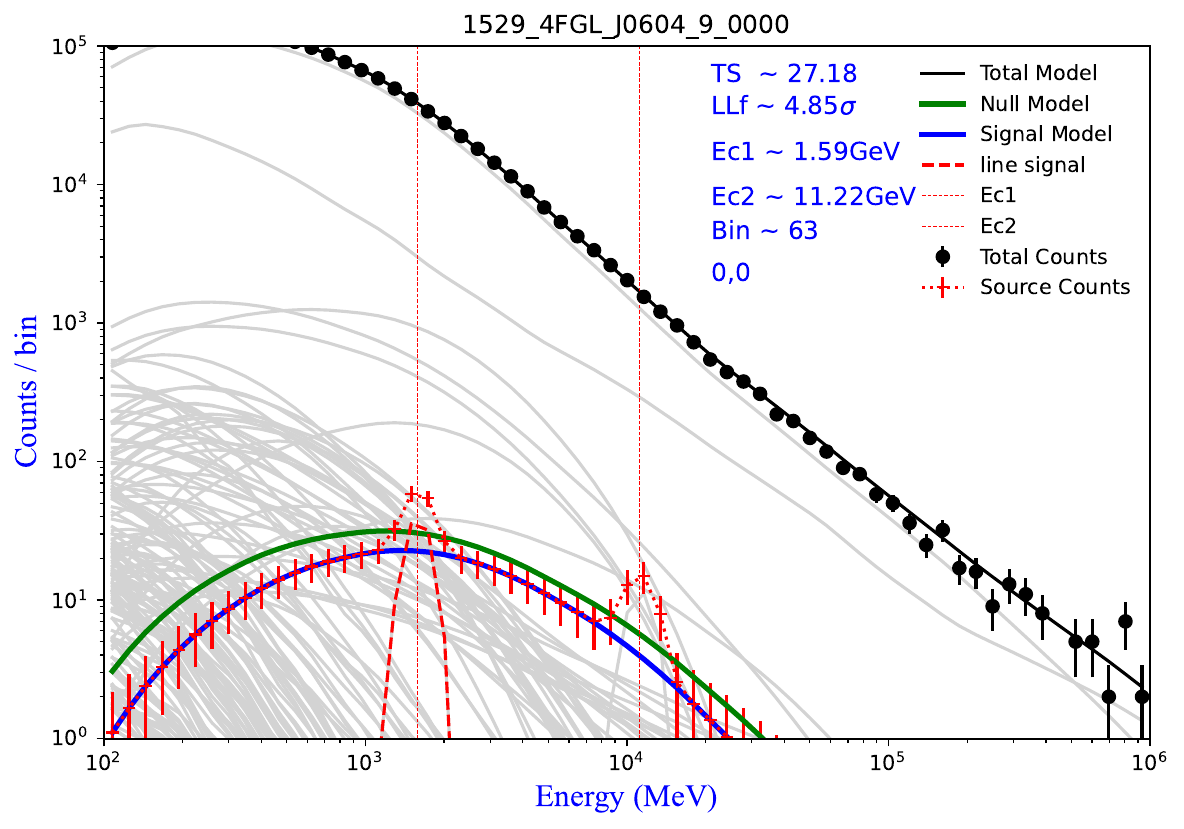}
    \includegraphics[width=0.32\linewidth]{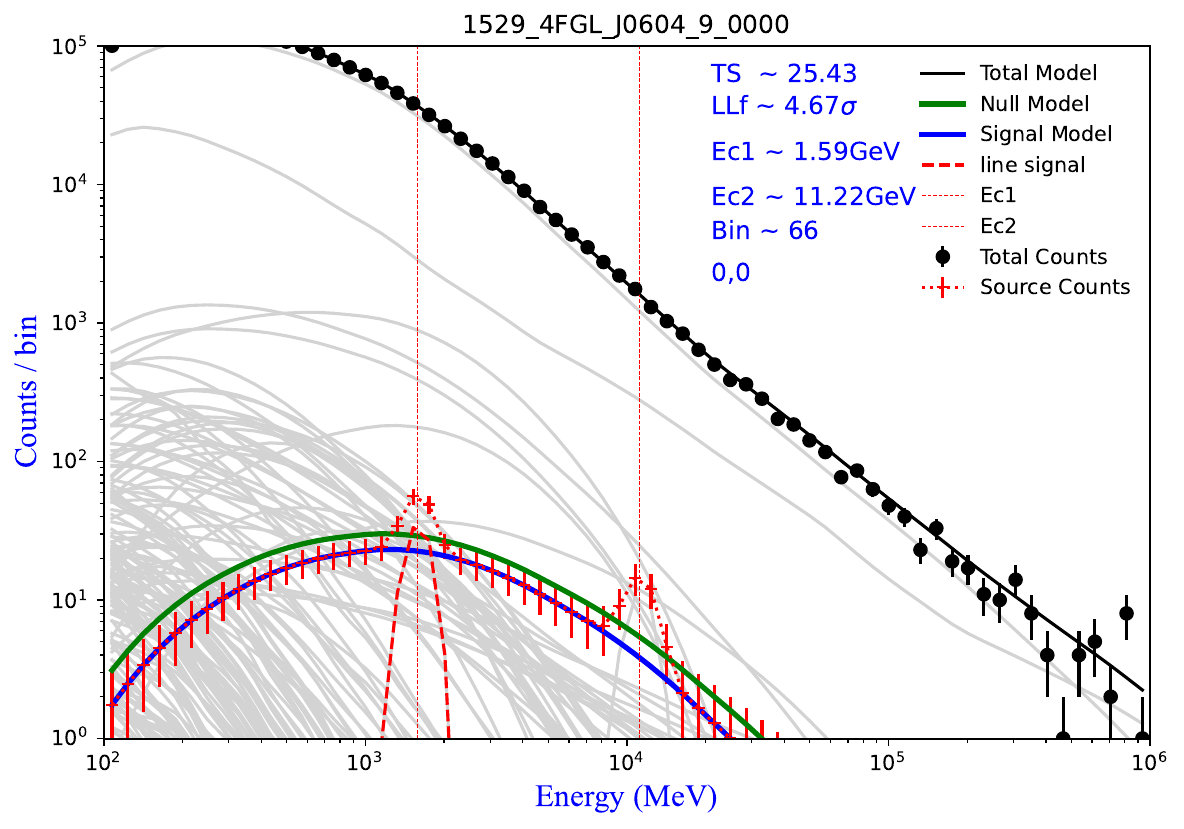}
    \caption{Model-fitted count spectra for the joint analysis of two excess signals in 4FGL J0604.9-0000 over 17.75 years. 
    The energy range from 100~MeV to 1~TeV is divided into 56 (left panel), 63 (middle panel), and 66 (right panel) logarithmically spaced bins. 
    Black dots and black line: data and total model; light grey lines: contributions from background sources; red dot-dashed line: the blazar's intrinsic source model; blue solid line: log-parabola continuum; red dashed lines: two Gaussian components representing the excesses at $\sim$1.5 GeV and $\sim$11 GeV; green solid line: null hypothesis model (continuum only). Vertical dashed lines mark the best-fit centroids of the Gaussians. The fit uses fixed widths ($E_1 \simeq 1.587$ GeV,  $\sigma_1 \simeq 0.145$ GeV, $E_2 \simeq 11.217$ GeV, $\sigma_2 \simeq 1.185$ GeV) based on instrumental energy resolution, and corresponds to the joint analysis with 2 dof (see Table~\ref{tab1}).}
    \label{fig_joint}
\end{figure*}

As shown in Figure~\ref{fig_joint}, the time-averaged spectrum of 4FGL~J0604.9-0000 over 17.75 years reveals two distinct excesses superimposed on the log-parabola continuum. The lower-energy excess is centered at $E_1 = 1.59\pm0.07$ GeV (for 63 bins) with a width fixed to the instrumental energy resolution ($\sigma_1 = 0.145$ GeV). The higher-energy excess is centered at $E_2 = 11.15\pm0.61$ GeV with a width $\sigma_2 = 1.185$ GeV, comparable to the LAT resolution at that energy. Fitting the data with a log-parabola plus two Gaussian components yields a test statistic $\mathrm{TS}=27.08$ (for 4 dof) for the joint analysis, corresponding to a local significance of $4.3\sigma$ (see Table~\ref{tab1}).

\begin{figure*}[htbp!]
    \centering
    \includegraphics[width=0.32\linewidth]{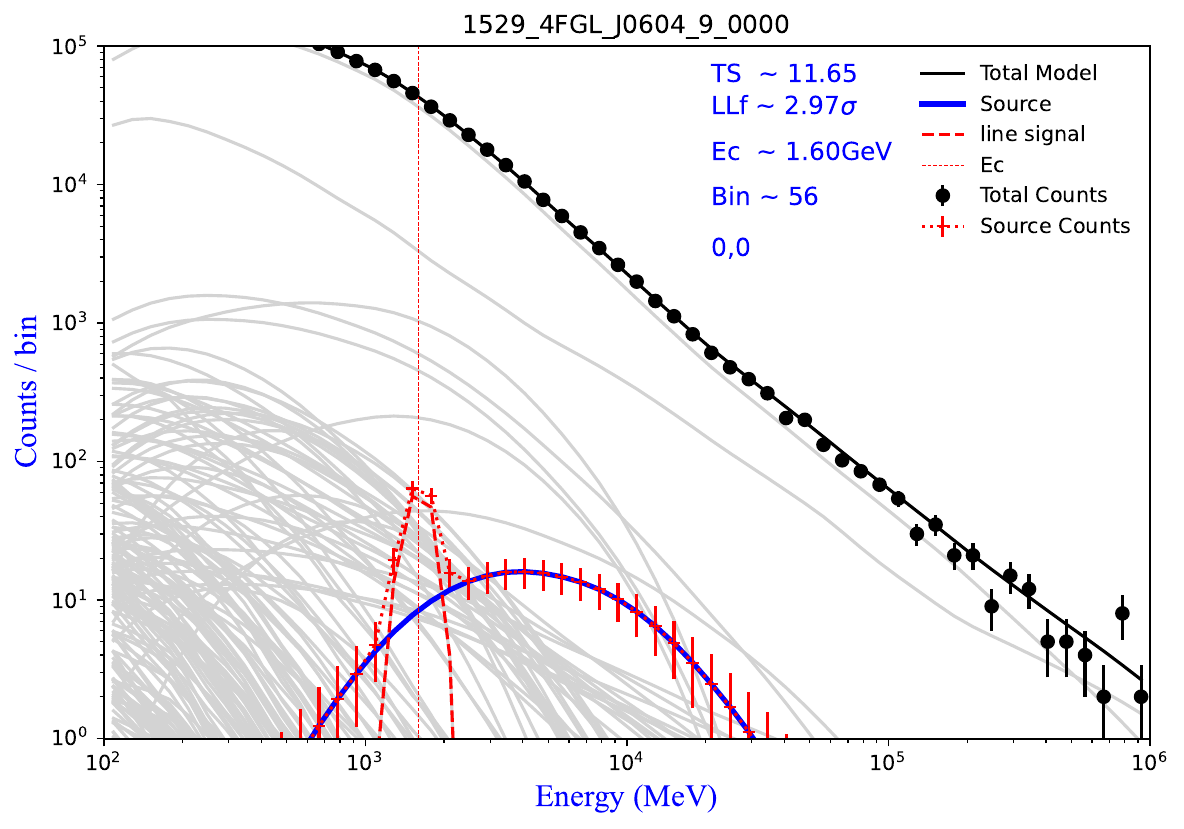}
    \includegraphics[width=0.32\linewidth]{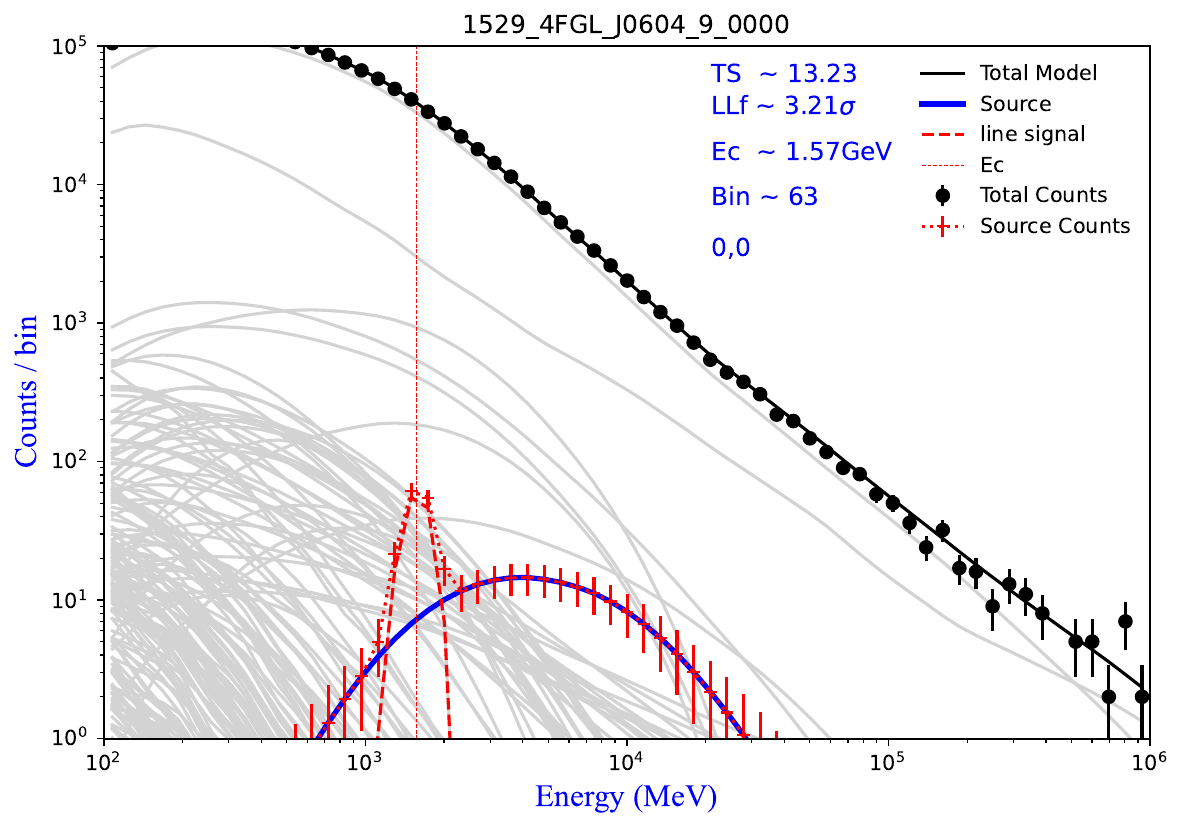}
    \includegraphics[width=0.32\linewidth]{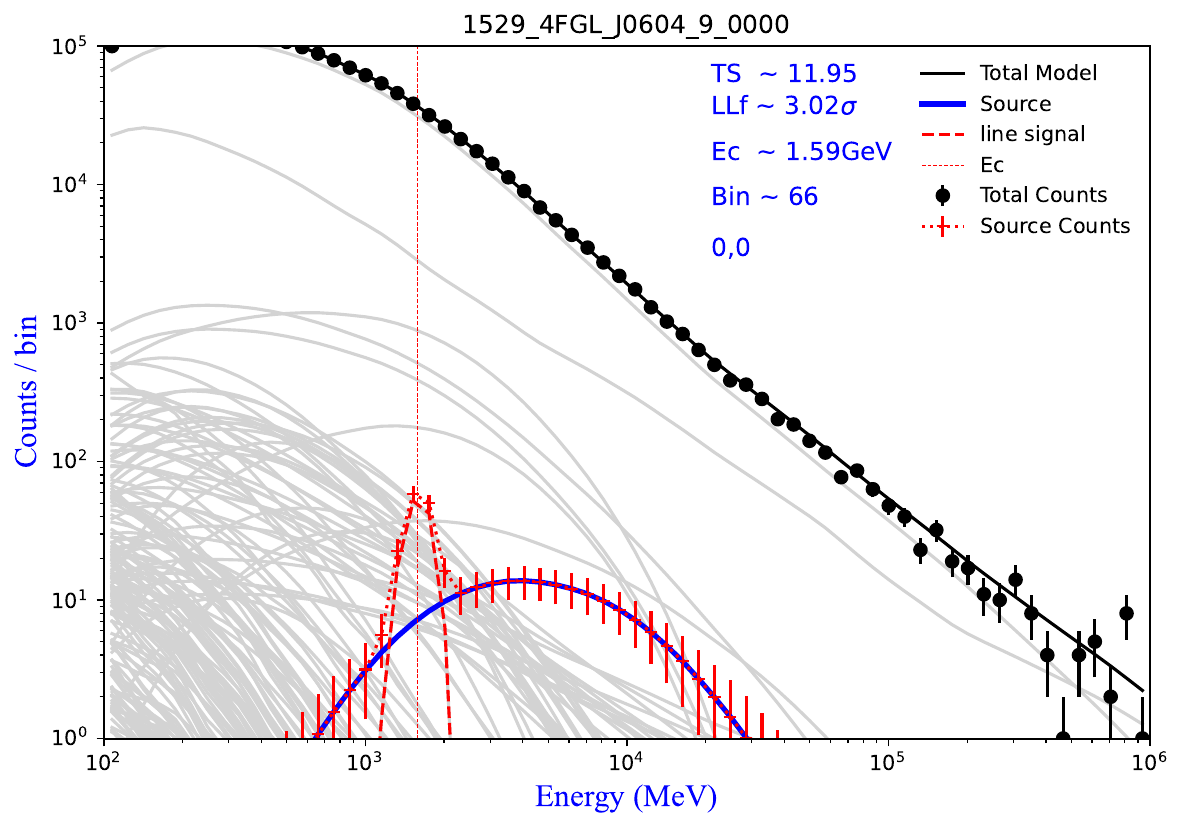}
    \caption{Same format as Figure~1, but showing only the single excess at $\sim$1.5 GeV (Signal~1). The Gaussian width is fixed to $\sigma_1 = 0.145$ GeV (2 dof fit). The panels correspond to 56, 63, and 66 logarithmically spaced energy bins from left to right. The local significance of this excess is $2.6\sigma$ (3 dof) or up to $3.5\sigma$ (1 dof) depending on the degrees of freedom (see Table~\ref{tab1}).}
    \label{fig_signal1}
\end{figure*}

\begin{figure*}[htbp!]
    \centering
    \includegraphics[width=0.32\linewidth]{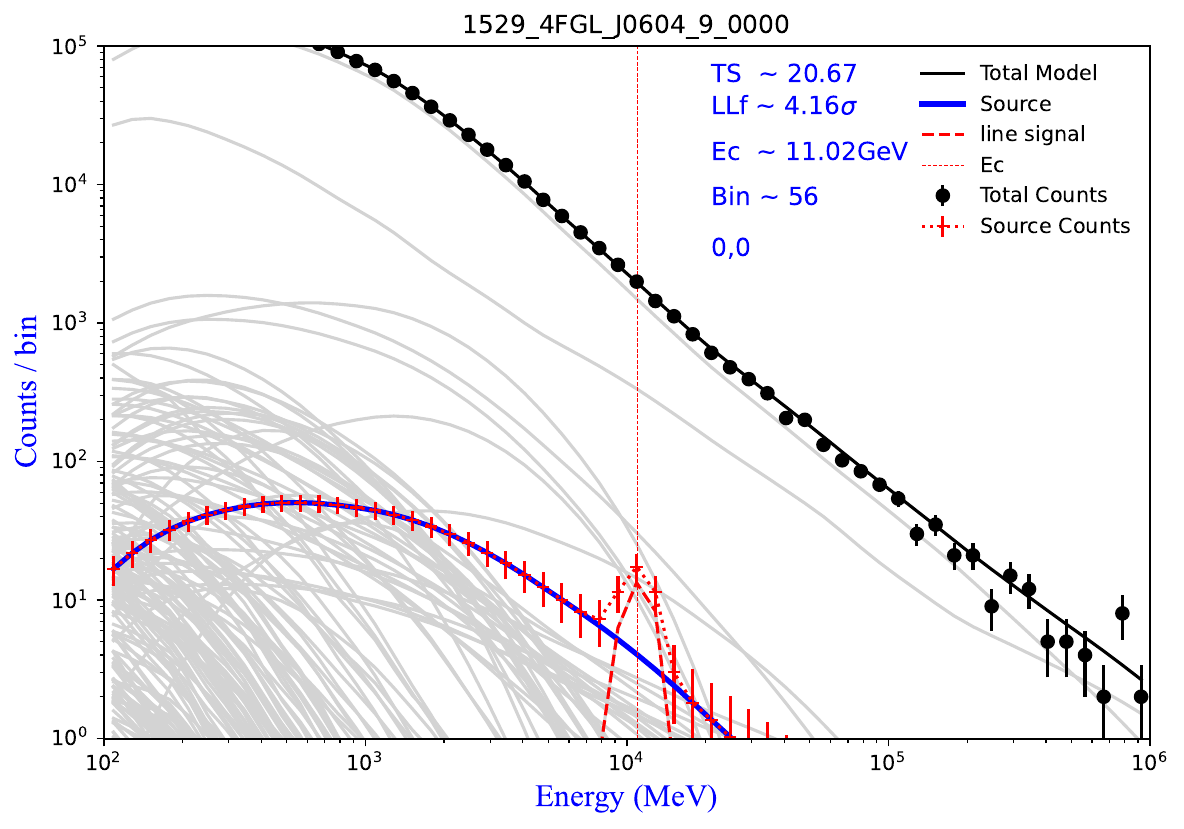}
    \includegraphics[width=0.32\linewidth]{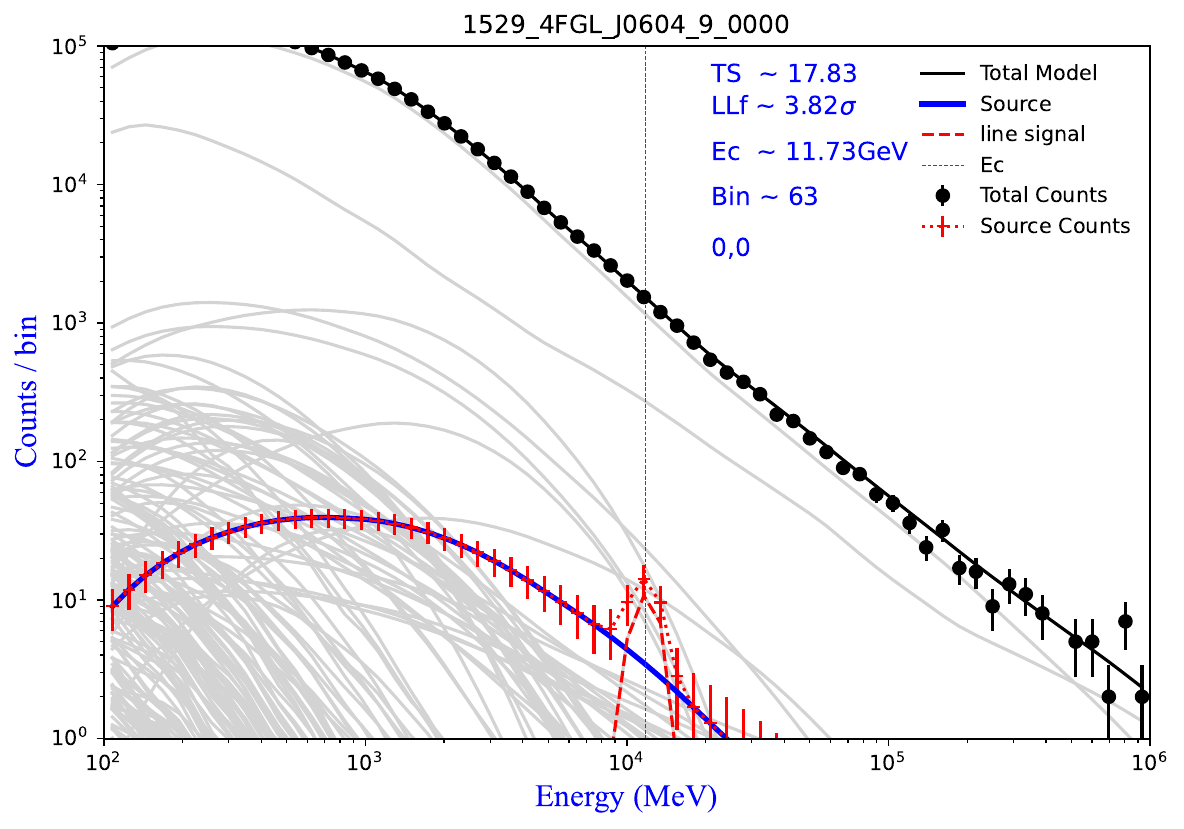}
    \includegraphics[width=0.32\linewidth]{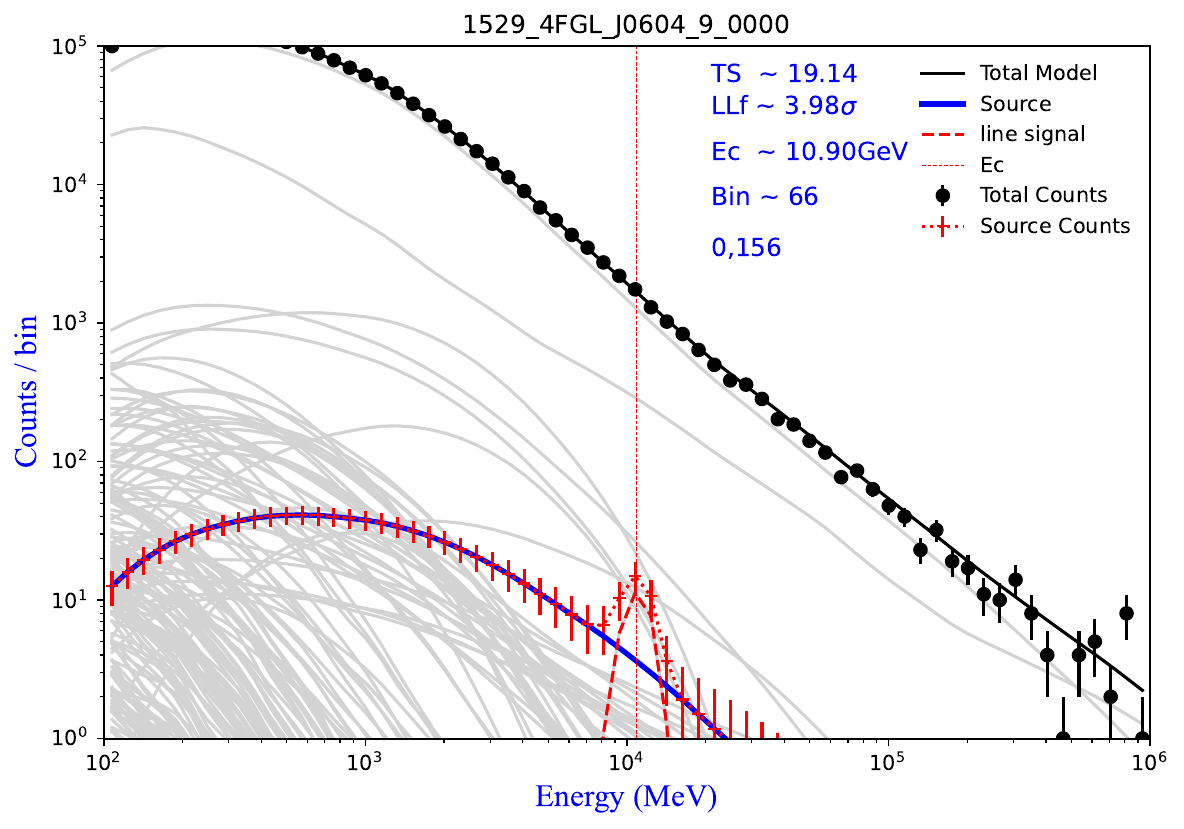}
    \caption{Same format as Figure~1, but showing only the single excess at $\sim$11 GeV (Signal~2). The Gaussian width is fixed to $\sigma_2 = 1.185$ GeV (2 dof fit). The panels show 56, 63, and 66 bins. The local significance of this excess is $3.7\sigma$ (3 dof) or up to $4.5\sigma$ (1 dof). The centroid energy varies slightly with binning ($E_2 \approx 10.9$–$11.7$ GeV); the joint analysis gives $E_2 = 11.15 \pm 0.61$ GeV (for 63 bins, 4dof, see Table~\ref{tab1}).}
    \label{fig_signal2}
\end{figure*}

When the line energies and widths are fixed to the best-fit values from the joint fit ($E_1=1.587$ GeV, $\sigma_1=0.145$ GeV, $E_2=11.217$ GeV, $\sigma_2=1.185$ GeV), the joint analysis with 2 dof gives $\mathrm{TS}=27.40$ (for 56 bins) and a local significance of $4.87\sigma$ (Table~\ref{tab1}). The single excess at $\sim$1.5 GeV (Signal~1) is detected with local significances ranging from $2.6\sigma$ (3 dof) to $3.5\sigma$ (1 dof) depending on the degrees of freedom (Figure~\ref{fig_signal1}). The $\sim$11 GeV excess (Signal~2) is detected with local significances from $3.7\sigma$ (3 dof) to $4.5\sigma$ (1 dof) (Figure~\ref{fig_signal2}). The centroid energy of the higher-energy excess shows mild dependence on the binning scheme ($E_2 \approx 10.9$–$11.7$ GeV), while the lower-energy excess remains stable (see Table~\ref{tab1}).

The observed energy ratio of approximately 1:7 is difficult to reproduce with conventional jet models, which typically produce smooth, featureless spectra \cite{2020ApJS..247...33A,2020ApJ...892..105A,2023arXiv230712546B,2026arXiv260222148B}. The sharpness of the excesses (widths comparable to or slightly larger than the LAT energy resolution) further disfavors standard astrophysical emission processes and points to a possible exotic origin.

\begin{table*}[ht!]
\caption{\label{tab1}
\textbf{Spectral Fitting Parameters for the {Excess} Signal.}
Column 1 provides the analytical information.
Column 2: The entire energy spectrum ranging from 100 MeV to 1 TeV is systematically divided into `N bins' equidistant energy intervals on a logarithmic scale. 
Columns 3 \& 4: Center energy and width of the fitting parameters for the first {excess} signal. 
Columns 5 \& 6: Center energy and width of the fitting parameters for the second {excess} signal. 
Columns 7 \& 8: Test statistic and corresponding local significance ($\sigma$, without correction for trials and look-elsewhere) for `N' degrees of freedom (`dof'). 
Here, the model-fitted count spectra (marked with *) are presented in Figures~\ref{fig_joint}-\ref{fig_signal2} to provide guidance regarding its form and content.
}
\begin{ruledtabular}
\begin{tabular}{lccccccc}
 &&\multicolumn{2}{c}{Signal 1\footnote{Parameters for $\sim$ 1.5 GeV signal.}}
  &\multicolumn{2}{c}{Signal 2\footnote{Parameters for $\sim$ 11 GeV signal.}}\\
 \cmidrule{3-4}\cmidrule{5-6} 
Analysis                   & Bin&  $E_{\rm 1}$ (GeV) & $\sigma_{\rm 1}$  (GeV) &  $E_{\rm 2}$ (GeV) & $\sigma_{\rm 2}$  (GeV) & TS  & $\sigma$ \\  
\hline
                           & 56 &	1.595 $\pm$ 0.077 	& 0.123$\pm$ 0.058 &	...     	     	&	...         	&	11.75 	&	2.64 	\\ 
Signal 1 (3 dof)   & 63 &    1.586 $\pm$ 0.065   & 0.147$\pm$ 0.062 &    ...     	           &   ...             &   13.27   &   2.87    \\ 
                           & 66 &	1.589 $\pm$ 0.074 	& 0.166$\pm$ 0.073 &	...     	     	&	...         	&	12.06 	&	2.69 	\\                            
\cdashline{1-8}[1pt/1pt]
                                         & 56* &	 1.598 $\pm$ 0.070 	 &	 0.145   	    &	...     	     	 &	...         	 &	 11.65 	 &	 2.97 	\\ 
Signal 1 (2 dof)                 & 63* &  1.574 $\pm$ 0.064 	&   0.145   	   &    ...     	     	&   ...         	&   13.23   &   3.21    \\ 
fixed $\sigma_{\rm 1}$     & 66* &	1.589 $\pm$ 0.066 	&	0.145   	   &	...     	     	&	...         	&	11.95 	&	3.02 	\\ 
\cdashline{1-8}[1pt/1pt]
                                        & 56  &	 1.587  	  	  	 &	0.145   	   &	...     	     	&	...         	&	11.63 	&	3.41 	\\   
Signal 1 (1 dof)                & 63  &.      1.587  	  	  	  &  0.145   	      &    ...     	           &   ...         	   &   13.26   &   3.64    \\ 
fixed $\sigma_{\rm 1}$, $E_{\rm 1}$  
                                       & 66  &	 1.587  	  	  	 &	0.145   	   &	...     	     	&	...         	&	11.95 	&	3.46 	\\ 
\hline
                           & 56 &	        ...         &	...            &	11.733 $\pm$ 6.137 	& 0.915 $\pm$ 6.137 &	19.22 	&	3.67 	\\ 
Signal 2 (3 dof)   & 63 &             ...        &	...           &         10.443 $\pm$ 0.937  & 2.182 $\pm$ 1.353 &   19.59   &   3.71    \\ 
                           & 66 &	        ...         &	...            &	10.910 $\pm$ 1.246 	& 0.457 $\pm$ 0.781 &	19.73 	&	3.73 	\\
\cdashline{1-8}[1pt/1pt]
                                          & 56* &	         ...         &	...     	   &	11.019 $\pm$ 0.627  & 1.185     	    &	20.67 	&	4.16 	\\ 
Signal 2 (2 dof)                 & 63* &   	    ...         &	...   	      &	   11.732 $\pm$ 0.609  & 1.185     	       &   17.82   &   3.82    \\ 
fixed $\sigma_{\rm 2}$     & 66* &	        ...         &	...     	  &	   10.897 $\pm$ 0.511  & 1.185     	       &   19.14   &   3.98 	\\ 
\cdashline{1-8}[1pt/1pt]
                                        & 56 &	        ...         &	...     	   &	11.217              & 1.185     	    &	20.58 	&	4.54 	\\ 
Signal 2 (1 dof)               & 63 &   	    ...        &	...   	      &	   11.217              & 1.185     	       &   18.90   &   4.35    \\ 
fixed $\sigma_{\rm 2}$, $E_{\rm 2}$   
                                       & 66 &	        ...         &	...     	   &	11.217              & 1.185     	    &	18.75   &   4.33 	\\ 
\hline
\cdashline{1-8}[1pt/1pt]
                                       & 56 &	1.604 $\pm$ 0.079 	&	0.145          &	11.022 $\pm$ 0.642 	& 1.185     	    &	27.53 	&	4.32	\\ 
Joint Analysis (4 dof)     & 63 &  1.587 $\pm$ 0.070   &      0.145          &    11.151 $\pm$ 0.610  & 1.185       	   &   27.08.       &   4.27    \\ 
fixed $\sigma_{\rm 1}$, $\sigma_{\rm 2}$ 
                                      & 66 &	1.592 $\pm$ 0.082 	&	0.145          &	10.916 $\pm$ 0.526 	& 1.185     	    &	26.32 	&	4.19	\\
\cdashline{1-8}[1pt/1pt]
                                      & 56* &	1.587            	&	0.145          &	11.217     	      	& 1.185     	    &	27.40 	&	4.87	\\
Joint Analysis (2 dof)     & 63* & 1.587               &   0.145             &	11.217     	        & 1.185     	    &   27.08        &   4.84	   \\
fixed $E_{\rm 1}$, $\sigma_{\rm 1}$, $E_{\rm 2}$, $\sigma_{\rm 2}$ 
                                      & 66* &	1.587            	&	0.145          &	11.217     	      	& 1.185     	    &	25.99 	&	4.73	\\
\end{tabular}
\end{ruledtabular}
\end{table*}

\section{\label{sec:discussion}Discussion}

The appearance of two narrow, sharpness excesses in a single AGN is reminiscent of the long-sought `smoking-gun' signature of dark matter (DM) annihilation \citep{1998APh.....9..137B,2012PDU.....1..194B}.
If interpreted as DM annihilation into $\gamma\gamma$ and $\gamma\gamma'$ (where $\gamma'$ is a dark photon), the higher-energy line at $\sim$11 GeV would correspond to the DM mass $m_{\chi}\approx 11$ GeV, while the lower-energy line at $\sim$1.5 GeV would arise from the $\gamma\gamma'$ channel. The required annihilation cross section is $\langle \sigma v\rangle_{\gamma\gamma} \sim 10^{-28}-10^{-27}~\mathrm{cm}^3\,\mathrm{s}^{-1}$, which is within the range of loop-suppressed annihilations for GeV-scale weakly interacting massive particles \citep{2012PDU.....1..194B}.

Intriguingly, a persistent gamma-ray line at $\sim$1.5 GeV has recently been reported in three AGNs (including blazars and a radio galaxy) by \citep{2026arXiv260400579K}, with a joint significance exceeding $5\sigma$. The energy of the lower excess in this source ($E_1 = 1.59\pm0.07$ GeV) is fully consistent with that feature. While \citep{2026arXiv260400579K} interpreted the 1.5 GeV line as a dark matter annihilation signal from a $\sim$1.5 GeV particle, our observation of an additional line at $\sim$11 GeV in the same source suggests a more complex scenario, possibly involving a heavier dark matter particle ($m_\chi \approx 11$ GeV) with two distinct annihilation channels ($\gamma\gamma$ and $\gamma\gamma'$). The $\sim$1.5 GeV line observed in other AGNs could then arise from the $\gamma\gamma'$ channel of an 11 GeV DM particle, rather than from a separate 1.5 GeV DM particle.

Another notable line candidate was reported at $\sim$43 GeV in stacked observations of nearby galaxy clusters \citep{2016PhRvD..93j3525L,2021ApJ...920....1S,2024arXiv240711737F}, with global significance $\sim$3-4$\sigma$ but later questioned as partly due to systematic effects \citep{2021ApJ...920....1S}. Our 11 GeV line does not correspond to that energy, but the existence of multiple line candidates across different astrophysical targets (Galactic center, galaxy clusters, AGN) at various energies ($\sim$1.5 GeV, $\sim$11 GeV, $\sim$43 GeV, $\sim$130 GeV) highlights the need for a unified understanding, either via a common dark matter particle with multiple final states or via diverse astrophysical processes. The double-line signature in a single object, however, provides stronger evidence against a purely astrophysical origin than isolated single lines.
More broadly, multi-mass or multi-line scenarios have been invoked to interpret such candidate complexes in other targets (e.g., stacked clusters, \citep{2026arXiv260617044K}), yet our single-source double excess offers a distinct and more constrained fingerprint.

Alternative explanations cannot be ruled out at present. The excesses could, in principle, originate from instrumental artifacts, but their stability over 17.75 years and the absence of similar features in nearby sources disfavor such an origin. Unresolved nearby point sources or complex Galactic diffuse emission are unlikely given the source's location at $|b|=10.3^\circ$ and the use of multiple background models. Hadronic processes in the jet, such as pion decay from proton-proton collisions, could produce gamma-ray bumps, but they typically yield broader features and are usually accompanied by neutrino emission not yet observed from this source \citep{2013ApJ...768...54B}. Photon-axion-like-particle oscillations in the jet's magnetic field might also produce spectral modulations \citep{2021JCAP...08..007Z}, but such models generally predict oscillatory patterns rather than two isolated narrow excesses.

If confirmed by future observations, the double-excess structure would represent the first detection of two distinct gamma-ray lines from a single AGN and would provide strong evidence for non-standard physics, either in the form of DM annihilation or exotic processes in the jet. The tentative nature of the signal -- owing to the moderate local significance and the absence of a full trial correction -- necessitates further scrutiny. Continued monitoring with Fermi-LAT, combined with observations from next-generation instruments such as the Very Large Area $\gamma$-ray Space Telescope (VLAST) \citep{2022AcASn..63...27F} and the Cherenkov Telescope Array \footnote{\url{https://www.cta-observatory.org/}} (CTA) \citep{2013APh....43....3A}, will be crucial to verify the reality of this feature and to discriminate among possible interpretations.

\section{\label{sec:conclusions}Conclusions}
In summary, we have reported a tentative double excess in the GeV gamma-ray spectrum of the blazar 4FGL~J0604.9-0000. While the statistical evidence is not yet conclusive, the unusual spectral morphology and the difficulty of accommodating it within conventional jet models make this source a compelling target for future deep observations. Confirmation of this signal would open a new window into the physics of relativistic jets and could provide the first indirect evidence for particle dark matter in an AGN environment.

\begin{acknowledgments}
  We thank the anonymous editor and referee for very constructive and helpful comments and suggestions, which greatly helped us to improve our paper.
  We thank Y. Yin, Y. G. Zheng, and Q. W. Wu for helpful discussions. 
  S. J. Kang acknowledges the support of the National Natural Science Foundation of China (Grant No.12163002) and the Liupanshui Science and Technology Development Project (Grant No. 52020-2024-PT-01) and the Discipline-Team of Liupanshui Normal University (LPSSY2025XKTD07).
\end{acknowledgments}

\section*{Software} 
    This work made use of the Fermi Science Tools (version 2.2.0). 
     
\section*{Data availability}
     The Fermi-LAT data underlying this article are publicly available from the Fermi Science Support Center (FSSC) and the High Energy Astrophysics Science Archive Research Center (HEASARC) at \url{https://fermi.gsfc.nasa.gov/ssc/}. Derived data products and analysis scripts are available from the corresponding author upon reasonable request.


\bibliographystyle{apsrev4-2}
\bibliography{apssamp}
 
\end{document}